\documentstyle[12pt]{article}
\begin{document}

\flushbottom
\newcommand\ie {{\it i.e. }}
\newcommand\eg {{\it e.g. }}
\newcommand\etc{{\it etc. }}
\newcommand\cf {{\it cf.  }}
\newcommand\viz{{\it viz. }}
\newcommand\grad{\nabla}
\newcommand\noi{\noindent}
\newcommand\seq{\;\;=3D\;\;}
\newcommand\barcaps{\cal}
\newcommand\jump{\vspace*{17pt}}
\newcommand\emptypage{~~~ \eject}
\setlength{\baselineskip}{17pt}
\def\be{\begin{eqnarray}}
\def\ee{\end{eqnarray}}
\newenvironment{draftequation}[1]{\be\label{#1}}{\ee}
\newcommand\bbe[1]{\begin{draftequation}{#1}}
\newcommand\eee{\end{draftequation}}
\def\ps{p\hspace{-0.075in}/}
\def\pis{\pi\hspace{-0.075in}/}
\def\half{{\textstyle{1 \over 2}}}
\def\ihalf{{\textstyle{i \over 2}}}
\def\D{{\cal D}}
\newcommand\art[1]{\cite{#1}}
\newcommand\ekv[1]{(\ref{#1})}

\def\aligneq#1#2{\cr\cr\noalign{\hbox to \hsize{#1\hfil}\vskip-3\baselineskip
    \hbox to \hsize{\hfil(\arabic{equation})}\addtocounter{equation}{1}
    \vskip2\baselineskip\llap{\hbox to .5cm{\hfil}}}\label{#2}}

\begin{flushright}
USITP-95-03\\
March 1995
\end{flushright}

\bigskip
\Large
\begin{center}
\bf{Hamiltonian BRST Quantization of the Conformal Spinning String}

\bigskip

\normalsize
by
\bigskip

P. Saltsidis\footnote{e-mail address:panos@vanosf.physto.se}\\
{\it ITP\\
University of Stockholm\\
Box 6730, Vanadisv\"agen 9\\
S-113 85 Stockholm\\
SWEDEN}\\
\end{center}
\vspace{2.0cm}
\normalsize
\bigskip
\bigskip

{\bf Abstract:}
The BRST quantization of the null spinning string for different number of
supersymmetries is
given.
A null spinning string with manifest space-time conformal invariance is
constructed.
Its BRST quantization gives  negative critical dimensions for $N\neq 0$ and
reproduces
previous results for $N=0$.

\eject

\begin{flushleft}
\large
{\bf Introduction}
\end{flushleft}
The characteristic scale of string theory is given by the string tension $T$.
At energies
of order of $\sqrt{T}$ or higher, string physics  truly distinguishes itself
 from point
particle physics. One may alternatively view the high energy limit as the zero
tension limit,
since only the energy measured in string units, $\frac{E}{\sqrt{T}}$ is
 relevant.
Strings in which
 the zero tension ($T\to 0$), has been taken are called null and they were
 first
considered by Schild \cite{sc}.
It has been shown in \cite{akul} that a null string corresponds to a collection
 of particles each
of which moves independently along a null geodesic. In this respect the zero
tension limit of a string
theory is  the analogue of massless particles.

The quantization of the null string was firstly discussed in \cite{liraspsr}
and then in \cite{gararual1}.
In \cite{liraspsr} was found that the quantization of the string does not give
 rise to critical dimensions,
while in \cite{gararual1} was pointed out that the issue depends crucially
on the choice of ordering,
and that normal ordering gives a critical dimension $d=26$ for the bosonic
null string. In \cite{jiulbs}, it
was shown that when $T\to 0$, the Weyl invariance  is substituted by conformal
invariance at the classical
level. In \cite{BigT} using general arguments it is shown that requiring this
symmetry to hold at the quantum level leads to
restrictions on the Hilbert space. The supersymmetric version of null string
is given in \cite{gararual1,ulbsgt1,ulbsgt2} and for general
where a null string with manifest space-time supersymmetry is given in
\cite{ulbsgt1} (null superstring) and
one with manifest world-sheet supersymmetry is given in
 \cite{gararual1,ulbsgt2}
(null spinning string).

In this paper we are going to investigate the conformal symmetry  of the null
spinning string at the
quantum level using  covariant methods and in particular the Hamiltonian BRST
method \cite{us}
using the ordering presented in
\cite{liraspsr} which we find is more natural for the tensionless string
 discussed in
\cite{BigT}.

The content of the paper is as follows:
In section 1 we give the BRST quantization of the usual null spinning string
 for
different numbers
of supersymmetries. There, using normal ordering we reproduce for $N=0$ and
 $N=2$
the results obtained
in \cite{gararual1}. Then we give the results of the BRST quantization
for the physical
ordering \cite{liraspsr,BigT,us}. In section 2 we construct a null spinning
 string
with manifest
conformal invariance. Using the same methods we find negative critical
 dimensions
for $N\neq 0$ and
$d=2$ when $N=0$, as found in \cite{us}.

\begin{flushleft}
\section{The null spinning string}
\end{flushleft}

The action for the null spinning string is given by \cite{ulbsgt2}
\bbe{action}
S=\int d^{2}\sigma [(V^{a}\partial_{a} X^{\mu}+i\Psi^{i\mu}\chi^{i})
  (V^{b}\partial_{b}
X^{\mu}+i\Psi^{j}_{\mu}\chi^{j})+i\Psi^{i\mu}V^{a}\partial_{a}\Psi_{\mu}^{i}],
\eee
where $\mu =0,\ldots ,d-1$ is a space-time index $d$ being the space-time
 dimension and
$a=0,1$ is a world sheet index. $V^{a}$ is a weight
$w=-\half$ contravariant 2-dimensional vector density,
$\Psi^{i\mu}$ is the fermionic partner of $X^{\mu}$, $i=1,\ldots ,N $,
$N$ being the
number of supersymmetries and $\chi^{i} $ is
the fermionic  partner of $V^{a}$. The fermions are densities of weight
 $w=-\frac{1}{4}$ to
ensure covariance. In the $T\to 0$ limit $\chi^{i} $ arises as
 $\chi^{i}=V^{a}\chi^{i}_{a}$
with $\chi^{i}_{a}$ the ordinary gravitino.
In (\ref{action}) and in all the following relations a summation over
 repeated $i$ indices
is assumed.

The action is invariant under the following local supersymmetry
 transformations
\be
\delta X^{\mu}&=&i\epsilon^{i}\Psi^{i\mu}\cr
\delta \Psi^{i\mu}&=&-\epsilon^{i}\partial X^{\mu}-
i\epsilon^{i}(\Psi^{j\mu}\chi^{j})-\half i\Psi^{i\mu}(\epsilon^{j}\chi^{j})\cr
\delta V^{a}&=&iV^{a}(\epsilon^{i}\chi^{i})\cr
\delta \chi^{i}&=&\nabla
\epsilon^{i}+\frac{3}{2}i(\epsilon^{j}\chi^{j})\chi^{i},\nonumber
\ee
where $\epsilon =\epsilon(\sigma)$ is a density of weight $\frac{1}{4}$ and
 we have
introduced the notation $\partial =V^{a}\partial_{a}$ and $ \nabla
 =V^{a}\nabla_{a}$. The
covariant derivative involves a connection about which it is sufficient
 to assume
$\nabla_{a}
V^{a}=0$,
which thus is the "metricity condition" of our theory.

The canonical momenta derived from the action (\ref{action}) are
\be
P_{\mu}&\equiv &\frac{\partial L}{\partial \dot{X}^{\mu}}=2V^{0}(V^{a}\partial
 _{a}X_{\mu}
+i\Psi_{\mu}^{i}V^{a}\chi^{i}_{a})\cr
\pi^{i}_{\mu}&\equiv &\frac{\partial L}
{\partial\dot{\Psi}^{i\mu}}=-iV^{0}\Psi^{i}_{\mu}\cr
P_{\chi}^{i,a}&\equiv &\frac{\partial L}{\partial \dot{\chi}^{i}_{a}}=0\cr
P_{Va}&\equiv &\frac{\partial L}{\partial\dot{V^{a}}}=0.\nonumber
\ee
They satisfy the usual commutation relations. The set of primary
constraints is given by
\be
D_{\mu}^{i}\equiv \pi^{i}_{\mu}+iV^{0}\Psi^{i}_{\mu}=0,
 \quad P_{\chi}^{i,a}=0, \quad
P_{Va}=0.\nonumber
\ee

{}From the general definition of the Hamiltonian we can write
\be
H=\int d^{2}\sigma [\dot{X}^{\mu}P_{\mu} +
\dot{\Psi}^{\mu}\pi_{\mu}-L+\lambda^{i}_{1a}P_{\chi}^{i,a}+
  \lambda_{2}^{a}P_{Va}+\lambda_{3}^{i\mu}D^{i}_{\mu}],\nonumber
\ee
where the $\lambda$'s are Lagrange multipliers. In order for the theory
to be consistent
all the constraints must hold for all times. This means that we have to
 require the
following consistency conditions
\be
\dot{P_{\chi}}^{i,a}(\sigma)=[P_{\chi}^{i,a}(\sigma),H]=0
\Rightarrow S^{i,-1}(\sigma)\equiv P^{\mu}\Psi^{i}_{\mu}(\sigma)=0\nonumber
\ee
which means that the condition $ S^{i,-1}(\sigma)=0$ is a secondary constraint.
 In exactly
the  same way we will have
\be
\dot{P_{V}}_{1}(\sigma)=0 &\Rightarrow & \phi ^{L}(\sigma)\equiv
(\frac{1}{V^{0}}X'_{\mu}P^{\mu} +i\Psi^{\mu}\Psi'{\mu})(\sigma)=0\cr
\dot{P_{V}}_{0}(\sigma)=0 &\Rightarrow &
\frac {1}{2{V^0}^3}P^{\mu}P_{\mu}-\frac{1}{{V^0}^2}
V^1 X'^{\mu}P_{\mu}-i\lambda^{i\mu}_{3}\Psi_{\mu}^{i}=0\cr
\dot {D}_{\mu}^{i}(\sigma)=0 &\Rightarrow & 2i\Psi^{i\mu}V^{1}\Psi '_{\mu}
-2iV^{0}\lambda^{i\mu}_{3}\Psi^{i}_{\mu}=0\nonumber
\ee
and using these three last relations
\be
\phi^{-1}(\sigma) \equiv P^{\mu}P_{\mu}(\sigma)=0.\nonumber
\ee

 It is not difficult to show now that ${P_{V}}_{a}(\sigma),
 {P_{\chi}}^{a}(\sigma),
S^{i,-1}(\sigma), \phi^{L}(\sigma) $ and $\phi^{-1}(\sigma)$
 constitute a complete set
of first class constraints and that the Hamiltonian is a linear combination of
these\footnote{This is in fact something that we might expected since the
 theory is
reparametrization invariant.}. In what follows we are going to gauge
 fix the field
$V^{0}(\sigma)$ equal to
$\half$ to have  a direct correspondence with previously obtained results. The
remaining constraints form
the following algebra
\bbe{CALG}
\left[\phi^L(\sigma ),\phi^{L}(\sigma ' )\right]_{D.B.}&=&
2\left[\phi^L(\sigma )
+\phi^L(\sigma ' )\right]\frac{d}{d\sigma}\delta (\sigma - \sigma ')\cr
\left[\phi^{L}(\sigma),S^{i,-1}(\sigma')\right]_{D.B.}&=&
2\left[S^{i,-1}(\sigma )+\half S^{i,-1}(\sigma ' )\right]
\frac{d}{d\sigma}\delta (\sigma - \sigma')\cr
\left\{S^{i,-1}(\sigma ),S^{j,-1}(\sigma ')\right\}_{D.B.}&=&
-\frac{i}{2}\left[\phi^{-1}(\sigma )
+\phi^{-1}(\sigma ' )\right]\delta (\sigma - \sigma ')\delta^{ij}.
\eee
All other brackets are zero. Note that we replaced the original
 Poisson brackets
with Dirac brackets to eliminate the second class constraints .  This means
that the canonical commutation relations will be
\bbe{comereal}
\left[X^{\mu}(\sigma ),P_{\nu}(\sigma ' )\right]_{D.B.}&=
&\delta^{\mu}_{\nu}\delta
  (\sigma -\sigma ')\cr
\left\{\Psi^{i\mu}(\sigma),\Psi^{j}_{\nu}(\sigma ')\right\}_{D.B.}
&=&-i\delta^{ij}\delta^{\mu}_{\nu}\delta (\sigma -\sigma ').
\eee

To quantize the system we have to replace Dirac brackets
by (anti-)commutators according to
$i\{\hspace{.1in}\}_{(D.B.)\pm}\to[\hspace{.1in}]_{\pm}\hspace{.08in}
(\hbar\equiv 1)$.  Then (\ref{comereal})  become
\be
[x^{\mu}_{m},p^{\nu}_{n}]=i\delta_{m+n} \eta^{\mu\nu},\qquad
[\psi^{i}_{\mu},\psi^{j}_{\nu}]=\delta^{ij}_{m+n}\eta_{\mu\nu},\nonumber
\ee
where we have used the Fourier modes of the operators $X^{\mu}$
 and $\Psi^{i}_{\mu}$.
We should also note here that the mode indices  of $\psi^{i\mu}_{m}$
 are integral or half
odd depending  on the sector.  In the Ramond sector they are integral
and in the
Neveu-Schwarz sector they are half odd.
 So in the quantum case the constraint algebra (\ref{CALG}) takes the form
\be
\left[ {\phi}^{-1}_{m},{\phi}^{L}_{n}\right]   &=&
                (m-n){\phi}^{-1}_{m+n}\label{excont}\\
  \left[ {\phi}^{L}_{m},{\phi}^{L}_{n}\right] &=&(m-n)
    {\phi}^{L}_{m+n}+({d}_{1}m^{3}+{d}_{2}m)\delta_{m+n}
   \label{d12} \\
  \left[{\phi}^{L}_{m} ,{S}^{i,-1}_{n}\right]   &=&
         (\frac{m}{2}-n){S}^{i,-1}_{m+n}\label{LS}\\
\left\{S^{i,-1}_{m},{S}^{j,-1}_{n}\right\}&=&
                \half{\phi}^{-1}_{m+n}\delta^{ij},\label{SS}
\ee
where we have used the Fourier modes of the constraints which read
\be
\phi^{-1}_{m}&=&\half \sum_{-\infty}^{+\infty}p_{m-k}\cdot p_{k}=
0\label{ccon1}\\
\phi^{L}_{m}&=&-i\sum_{-\infty}^{+\infty}[kp_{m-k}\cdot x_{k}+\half
k\psi^{i}_{m-k}\cdot\psi^{i} _{k}]= 0\label{ccon2}\\
S^{i,-1}_{m}&=&\half \sum_{-\infty}^{+\infty}p_{m-k}\cdot \psi^{i}_{k}=
0\label{ccon3}.
\ee
Note that we have included central extensions in the constraint algebra due to
ordering ambiguities. The particular form of the central extensions is limited
by the Jacobi identities.

With the structure constants at hand we may write down the  Hamiltonian
BRST charge $\cal Q$
\bbe{Q2}
{\cal Q}&=& \sum_{k}
(\phi_{-k}^{-1}c_{k}^{-1}+
\phi^{L}_{-k}c_{k}^{L}+S_{-k}^{i,-1}\gamma_{k}^{i,-1})\nonumber\\
&-&
\sum_{k,l} [(k-l)c_{-k}^{-1}c_{-l}^{L}
b_{k+l}^{-1}+\half
(k-l)c_{-k}^{L}c_{-l}^{L}b_{k+l}^{L} -
(\frac{k}{2}-l)c_{-k}^{L}\gamma_{-l}^{i,-1}\beta_{k+l}^{i,-1}+\nonumber\\
 & & \frac{1}{4} \gamma_{-k}^{i,-1}\gamma_{-l}^{i,-1}b_{k+l}^{-1} ].
\eee
Here we have introduced the (anti-) ghosts $(b^{\alpha},\beta^{i,-1})$,
$c^{\alpha},\gamma^{i,-1}$, corresponding to the constraints
 $\phi^{\alpha}, S^{i,-1}$,
fulfilling the canonical relations $(\alpha=-1,L)$
\be
\left\{ c^{\alpha}_{m},b^{\beta}_{n}\right\}&=&\delta^{\alpha\beta}
  \delta_{mn}\nonumber\\
\left[ \gamma^{i,-1}_{m},\beta^{j,-1}_{n}\right]&=&\delta^{ij}
  \delta_{mn}.\nonumber
\ee
The couplings are determined by the structure constants $f^{IJK}$ of the
algebra\ekv{CALG} according to the general prescription of \art{FRAD}:
\bbe{eqnhere}
 {\cal Q}=\phi^I c^J - \half (-1)^{n_{J}} f^I_{JK}b_Ic^Kc^J,
\eee
where
\be
n_{J}=\left\{
\begin{array}{ll}0 & \mbox{for $\phi^J$ bosonic}\\
1 & \mbox{for $\phi^J$ fermionic}\nonumber
\end{array}
\right.
\ee
and the index $I$ runs over all possible constraints.

The classical nilpotency, ${\cal Q}^2=0$, is guaranteed by
construction.
To check the nilpotency of the quantum ${\cal Q}$ we use the following trick.

We begin by defining the extended constraints $\tilde{\phi}^{I}_{n}$ by
the equation
\bbe{tildephi}
\tilde{\phi}^{I}_{n}\equiv \{ b^{I}_{n},{\cal Q}\}.
\eee
Using (\ref{Q2})  we find that the extended
constraints are given by the following relations
\be
\tilde{\phi}^{-1}_{m}&=&:\!\phi^{-1}_{m}\! :- \sum_{-\infty}^{+\infty}:
(m-k)c^{L}_{-k}b^{-1}_{m+k}:
\label{excon1}  \\
\tilde{\phi}^{L}_{m}&=&:\! \phi^{L}_{m}\! :+ \sum_{-\infty}^{+\infty}:
[(k-m)c^{-1}_{-k}b^{-1}_{m+k}+(k-m)c^{L}_{-k}b^{L}_{m+k}+\nonumber \\
& &+(\frac{m}{2}-k)\gamma^{i,-1}_{-k}\beta^{i,-1}_{m+k}]:-\alpha_{L}\delta_{m}
\label{excon4}\\
\tilde{S}^{i,-1}_{m}&=&:\! S^{i,-1}_{m}\! :+ \sum_{-\infty}^{+\infty}:
[(\frac{k}{2}-m)c^{L}_{-k}\beta^{i,-1}_{m+k}
-\half \gamma^{i,-1}_{-k}b^{-1}_{m+k}]:,
\label{exconS-1}
\ee
where  : : is the  ordering for which ${\cal Q}\left|
phys\right\rangle =0$.

It is not difficult to check that the extended constraints satisfy the same
 algebra
as the original constraints so their algebra is  the tilded version of
(\ref{excon1})-(\ref{exconS-1}).

We can now calculate the BRST anomaly using a method described in
\cite{MarnABRST,ISBE}. There it is shown that

\be\label{Q}
{\cal Q}^{2}=\half \sum_{i,j}{\tilde {d}}^{IJ}_{m}c^{I}_{m}c^{J}_{-m},
\ee
where ${\tilde {d}}^{IJ}$ are the central extensions of the extended
 constraints
algebra.

This means that
 \be\label{as}
{\cal Q}^{2}&=&\tilde{d}_{1} \sum_{m}
\frac{m^{3}}{2}c^{L}_{m}c^{L}_{-m}+\tilde{d}_{2}
\sum_{m} \frac{m}{2}c^{L}_{m}c^{L}_{-m}.
\ee
The exact values of $\tilde{d}_{f}, f=1,2 $ depend on the
vacuum and ordering we have used. The simplest and safest method to determine
 these
constants is to calculate the vacuum expectation value of the commutators
(\ref{excon1})-(\ref{exconS-1}) for the extended constraints.

We assume first that every operator with positive index annihilates the
 vacuum. This means
that $ \forall m>0 $
\be
\tilde{\phi}^{-1}_{m}\left| 0\right\rangle  =
\tilde{\phi}^{L}_{m}\left|0\right\rangle =
\tilde{S}^{i,-1}_{m}\left| 0\right\rangle = 0.\nonumber
\ee
The expectation value of the commutator (\ref{d12}) is
\bbe{fcon1}
\left\langle 0\right|[\tilde{\phi}^{L}_{m},\tilde{\phi}^{L}_{-m}]\left|
  0\right\rangle =2m\left\langle 0\right|\tilde{\phi}^{L}_{0}\left|
  0\right\rangle +\tilde{d}_{1}m^{3}+\tilde{d}_{2}m.\nonumber
  \eee

In the R-sector for $m=1$ we will have
\be
2\left\langle 0\right|\tilde{\phi}^{L}_{0}\left|
  0\right\rangle +\tilde{d}_{1}+\tilde{d}_{2}=
\left\langle 0\right|\tilde{\phi}^{L}_{1}\tilde{\phi}^{L}_{-1}\left|
  0\right\rangle =\frac{Nd}{8}-4+\frac{3}{4}N.\nonumber
\ee
For $m=2$
\be
4\left\langle 0\right|\tilde{\phi}^{L}_{0}\left|
  0\right\rangle +8\tilde{d}_{1}+2\tilde{d}_{2}=
\left\langle 0\right|\tilde{\phi}^{L}_{2}\tilde{\phi}^{L}_{-2}\left|
  0\right\rangle =d+\frac{Nd}{2}-34+7N.\nonumber
\ee

{}From the last two equations we can write
\be
\tilde{d}_{1}=\frac{1}{24}(4d+Nd-104+22N)\nonumber\\
\tilde{d}_{2}=\frac{1}{24}(-4d+2Nd+8-4N-48\alpha^{R}_{L}),\nonumber
\ee
where $\alpha^{R}_{L}=\left\langle 0\right|\tilde{\phi}^{L}_{0}\left|
  0\right\rangle $. As we can see from (\ref{as}) the BRST charge can be
 nilpotent
only if $\tilde{d}_{1}=\tilde{d}_{2}=0$ which in turn means that the
 following relations
have to be satisfied
\be
d=\frac{104-22N}{4+N},\quad \alpha^{R}_{L}=-\frac {N^{2}-6N+8}{4+N}.\nonumber
\ee

So in the Ramond sector we will have positive critical space-time dimensions
 for the
following numbers of supersymmetries
\be
\begin{tabular}{|c||c|c|}
\hline
\multicolumn{3}{|c|}{Ramond Sector}\\ \hline
\multicolumn{1}{|c||}{N}&\multicolumn{1}{|c|}{Critical
dimension}&\multicolumn{1}{|c|}{$\alpha^{R}_{L}$}\\
\hline\hline
0 & 26 & -2\\ \hline
2 & 10 & 0 \\ \hline
4 & 2  & 0 \\ \hline
\end{tabular}\nonumber
\ee

In exactly the same manner we can find that in the NS-sector we will have
 the following
relations
\be
d=\frac{104-22N}{4+N}, \quad \alpha^{NS}_{L}=\frac
 {N^{2}-16}{2(4+N)}.\nonumber
\ee

So in the Neveu-Schwarz sector we will have the following table

\be
\begin{tabular}{|c||c|c|}
\hline
\multicolumn{3}{|c|}{Neveu-Schwarz Sector}\\ \hline
\multicolumn{1}{|c||}{N}&\multicolumn{1}{|c|}{Critical
dimension}&\multicolumn{1}{|c|}{$\alpha^{NS}_{L}$}\\
\hline\hline
0 & 26 & -2\\ \hline
2 & 10 & -1 \\ \hline
4 & 2  & 0 \\ \hline
\end{tabular}\nonumber
\ee
Note that the results for $N=0,2$ are those obtained in \cite{gararual1}.
 We   should
also remark  that the $N=4$ case  gives the same critical
dimension, $d=2$, with the $N=2$ formulation of the usual $T\neq 0$ spinning
 string.

According to arguments presented in \cite{BigT} however the  vacuum suitable
 for
tensionless is not
the one annihilated by the positive modes of the operators but the one
 annihilated
by the momenta
\bbe{FullVacCond3}
  p^{\mu}_{m}|0\rangle =0 \quad \forall m.\nonumber
\eee
Following the prescription of \cite{MarnBRST,us}, we will take
the {\em ket } states to be built from our vacuum of choice, $|0\rangle_{p}$,
and the {\em bra } states to be built from $\mbox{}_{x}\langle 0|$ satisfying
$\mbox{}_{x}\langle 0|0\rangle_{p}=1$. For this vacuum and from the requirement
that the BRST charge (\ref{Q}) should annihilate the vacuum, we can
 find requirements for
the ghost part of the vacuum. Doing this we find that the vacuum has to
 satisfy the
following conditions $\forall m,i$
\be
 p^{\mu}_m|0\rangle  =b^{-1}_m|0\rangle = \beta^{i,-1}_m|0\rangle=0\nonumber\\
  \langle 0|x^{\mu}_m  =\langle 0|c^{-1}_m=
 \langle 0|\gamma^{i,-1}_m=0.\nonumber
\ee
As can be seen from the previous relations the
 condition $p^{\mu}_m|0\rangle  =0$ does not
specify how the $\psi^{i\mu}_m$ operators will act on this vacuum. We
do not like to have a condition like $\psi^{i\mu}_m|0\rangle
=0$,$\forall m$ on the vacuum since this would destroy all the fermionic
creation
operators.
Choosing
$\psi^{i\mu}_m|0\rangle =0$,$\forall m>0$ will give us the following conditions
\bbe{F2}
  \psi^{i\mu}_m|0\rangle  =c^{L}_m|0\rangle = b^{L}_m|0\rangle=0\nonumber\\
  \langle 0|\psi^{i\mu}_{-m}  =\langle 0|c^{L}_{-m}=
\langle 0|b^{L}_{-m}=0.\nonumber
\eee
With  this choice of vacuum it is not difficult to find that the
 commutator (\ref{ccon2})
will give for $N\neq 0$\footnote{For $N=0$ it can be shown that
$\tilde{d}_{1}=\tilde{d}_{2}=0$ and thus that
${\cal Q}^{2}=0$, independent of the dimension $d$.}

\be
\tilde{d}_{1}&=&\frac{1}{24}(Nd-52)\nonumber\\
\tilde{d}^{R}_{2}&=&\frac{1}{24}(2Nd+4-48\alpha^{R}_{L}),\quad  \mbox{for the
R-sector,}\nonumber\\
\tilde{d}^{LS}_{2}&=&\frac{1}{24}(-Nd+4-48\alpha^{NS}_{L}),\quad \mbox{for the
NS-sector}.\nonumber
\ee
So in this case we will have the following
\be
\begin{tabular}{|c||c|c|c|}
\hline
\multicolumn{4}{|c|}{Physical Ordering}\\ \hline
\multicolumn{1}{|c||}{N}&\multicolumn{1}{|c|}{Critical
dimension}&\multicolumn{1}{|c|}{$\alpha^{R}_{L}$}&
\multicolumn{1}{|c|}{$\alpha^{NS}_{L}$}\\
\hline\hline
1 & 52 & $\frac{9}{4}$ & -1\\ \hline
2 & 26 & $\frac{9}{4}$ & -1\\ \hline
4 & 13 & $\frac{9}{4}$ & -1\\ \hline
13 & 4 & $\frac{9}{4}$ & -1\\ \hline
26 & 2 & $\frac{9}{4}$ & -1\\ \hline
52 & 1 & $\frac{9}{4}$ & -1\\ \hline
\end{tabular}\nonumber
\ee

\begin{flushleft}
\section{The conformal spinning string}
\end{flushleft}

Let us return to the action of the null spinning string (\ref{action}).
 It is not difficult
to check that the action is invariant under the group of space-time conformal
transformations
\begin{flushleft}(i) Lorentz transformations
\end{flushleft}
\bbe{ad}
\delta_{\omega} X^{\mu}&=&\omega^{\mu}_{\nu}X^{\nu}\cr
\delta_{\omega} \Psi^{i\mu}&=&\omega^{\mu}_{\nu}\Psi^{i\nu}\cr
\delta_{\omega} V^{a}&=&0\cr
\delta_{\omega} \chi^{i}_{a}&=&0\nonumber
\aligneq{(ii) Translations}{ADD}
\delta_l X^M &=& l^{\mu}\cr
\delta_l \Psi^{i\mu} &=&0\cr
\delta_l V^{a} &=& 0\cr
\delta_l \chi^{i}_{a} &=& 0\nonumber
\aligneq{(iii) Special conformal transformations}{ADD1}
\delta_{b} X^{\mu}&=&(b\cdot X)X^{\mu}-\half X^{2}b^{\mu}\cr
\delta_{b} \Psi^{i\mu}&=&\half (b\cdot X)\Psi^{i\mu}\cr
\delta_{b} V^{a}&=&-(b\cdot X)V^{a}\cr
\delta_{b} \chi^{i}_{a}&=&\half (b\cdot X)\chi^{i}_{a}\nonumber
\aligneq{(iv) Dilatations}{ADD2}
\delta_{a} X^{\mu}&=&aX^{\mu}\cr
\delta_{a} \Psi^{i\mu}&=&\frac{a}{2} \Psi^{i\mu}\cr
\delta_{a} V^{a}&=&a V^{a}\cr
\delta_{a} \chi^{i}_{a}&=&\frac{a}{2}\chi^{i}_{a}\nonumber
\eee
 The isomorphism $C_{d-1,1}\simeq O(d,2)$ for $d\geq 3$ makes it possible
 to construct a
theory in two extra dimensions such that the previous model corresponds
 to a particular gauge
fixing of the latter and the conformal symmetry is manifest and linearly
 realized
\cite{akul,ISBE,us}. This {\em conformal} spinning string action can
 be given by
\bbe{action2}
S&=&\int d^{2}\sigma \{[V^{a}(\partial_{a}+W_{a}) X^{A}+i\Psi^{iA}\chi^{i}]
  [V^{b}(\partial_{b}+W_{b})
X_{A}+i\Psi^{j}_{A}\chi^{j}]+\nonumber\\
& &i\Psi^{iA}V^{a}\partial_{a}\Psi_{A}^{i}+
iA^{ij}\Psi^{iA}\Psi^{j}_{A}+\Phi X^{A}X_{A}\},\nonumber
\eee

where $A=0,\ldots,d+1$ and the new metric has the form

\bbe{METR}
\eta _{AB} = \left( \matrix{\eta_{\mu\nu}\quad  \hfill 0 \quad 0  \cr
 0...0 \quad \hfill 1 \quad 0 \cr 0...0 \quad \hfill 0 -1 \cr}
\right).\nonumber
\eee
$W_{a}$ is the gauge field for scale transformations, $A^{ij}$ and $\Phi $
 are Lagrange
multiplier fields.

We can check that by imposing two gauge
 fixing conditions $P^{+}=0$, $X^{+}=1$ the
generators of the Lorentz transformations in the extended
 space become the generators of
the conformal group in the original space. Thus rotations in the
 extended space correspond
to conformal transformations in the original space.

Going to the Hamiltonian formulation we  find in exactly the same manner
 that the
Hamiltonian is again a linear combination of the constraints. In addition
 to the original
constraints (\ref{ccon1})-(\ref{ccon3}) we will have four new ones
 which in Fourier modes
can be written as follows
\be
\phi^{0}_{m}&=&\half \sum_{-\infty}^{+\infty}p_{m-k}\cdot x_{k}=
0\label{ccon4}\\
\phi^{1}_{m}&=&\half \sum_{-\infty}^{+\infty}x_{m-k}\cdot x_{k}=
0\label{ccon5}\\
S^{i,1}_{m}&=&\half \sum_{-\infty}^{+\infty}x_{m-k}\cdot \psi^{i}_{k}=
0\label{ccon6}\\
\phi^{ij}_{m}&=&\half \sum_{-\infty}^{+\infty}\psi^{i}_{m-k}\cdot \psi^{j}_{k}=
0\label{ccon7}.
\ee
where $i,j=1,\ldots,N$ and $i<j$ in the last constraint. Note that if
 one discards
$\phi^{L}_{m}$, the remaining constraints will  describe the conformal spinning
 particle
presented  in \cite{mart}.

The constraint algebra with the central extensions included will be given by
\be
\left[ {\phi}^{L}_{m},{\phi}^{ij}_{n}\right] &=&-n
           {\phi}^{ij}_{m+n}\label{dlij} \nonumber\\
\left[{S}^{i,-1}_{m} ,{\phi}^{\alpha\beta}_{n}\right]   &=&
           \frac{1}{2}{S}^{\beta,-1}_{m+n}\delta^{i\alpha}
           -\frac{1}{2}{S}^{\alpha,-1}_{m+n}\delta^{i\beta}
\label{dsij}\nonumber\\
\left[{\phi}^{ij}_{m} ,{\phi}^{\alpha\beta}_{n}\right]   &=&
          \frac{1}{2}{\phi}^{i\beta}_{m+n}\delta^{\alpha j}
          -\frac{1}{2}{\phi}^{\alpha j}_{m+n}\delta^{i\beta}+
          \frac{1}{2}{\phi}^{\beta j}_{m+n}\delta^{i\alpha }
         -\frac{1}{2}{\phi}^{i\alpha }_{m+n}\delta^{j\beta}
\label{dijab}\nonumber\\
\left[  {\phi}^{1}_{m},{\phi}^{-1}_{n}\right] &=&2i
          {\phi}^{0}_{m+n}+2(i{d}_{4}+i{d}_{3}m)\delta_{m+n}
          \label{d1-1} \nonumber\\
\left[ {\phi}^{1}_{m},{\phi}^{L}_{n}\right]   &=&
          (m+n){\phi}^{1}_{m+n}\nonumber\\
\left[{\phi}^{1}_{m} ,{S}^{i,-1}_{n}\right]   &=&
          i{S}^{i,1}_{m+n}\label{d1s}\nonumber\\
\left[ {\phi}^{-1}_{m},{\phi}^{0}_{n}\right]  &=&
         -i{\phi}^{-1}_{m+n}\nonumber\\
\left[ {\phi}^{0}_{m},{\phi}^{L}_{n}\right] &=&m
         {\phi}^{0}_{m+n}+({d}_{3}m^{2}+{d}_{4}m)\delta_{m+n}
         \label{d34} \nonumber\\
\left[{\phi}^{0}_{m} ,{S}^{i,-1}_{n}\right]   &=&
         \frac{i}{2} {S}^{i,-1}_{m+n}\nonumber\\
\left[ {\phi}^{1}_{m},{\phi}^{0}_{n}\right]  &=&
          i{\phi}^{1}_{m+n}\nonumber\nonumber\\
\left[{\phi}^{-1}_{m} ,{S}^{i,1}_{n}\right]   &=&
         -i{S}^{i,-1}_{m+n}\label{d-1s}\nonumber\\
\left[{\phi}^{L}_{m} ,{S}^{i,1}_{n}\right]   &=&
         -(\frac{m}{2}+n){S}^{i,1}_{m+n}\label{dls1}\nonumber\\
\left\{S^{i,-1}_{m},{S}^{j,1}_{n}\right\}&=&
         \half{\phi}^{0}_{m+n}\delta^{ij}
         -\frac{i}{2}{\phi}^{ij}_{m+n}
         +(\frac{{d}_{4}}{2}-m{d}_{3})\delta^{ij}\delta_{m+n}
\label{ds1s-1}\nonumber\\
\left[{S}^{i,-1}_{m} ,{\phi}^{\alpha\beta}_{n}\right]   &=&
           \frac{1}{2}{S}^{\beta,1}_{m+n}\delta^{i\alpha}
           -\frac{1}{2}{S}^{\alpha,1}_{m+n}\delta^{i\beta}
\label{ds-1ij}\nonumber\\
\left[{\phi}^{0}_{m} ,{S}^{i,1}_{n}\right]   &=&
         -\frac{i}{2} {S}^{i,1}_{m+n}\label{d0s}\nonumber\\
\left\{S^{i,1}_{m},{S}^{j,1}_{n}\right\}&=&
                \half{\phi}^{1}_{m+n}\delta^{ij}\label{ds1s1}
\nonumber\\
\left[ {\phi}^{0}_{m},{\phi}^{0}_{n}\right] &=&
        -im{d}_{3}\delta_{m+n}.
\ee
All the other commutators except (\ref{excont})-(\ref{SS}) vanish.

The structure constants obtained here can be inserted into the
relation (\ref{eqnhere})
to find the BRST charge which is rather complicated this time
due to the large number of
constraints
\bbe{Q3}
{\cal Q}&=& \sum_{k}(\phi_{-k}^{1}c_{k}^{1}+
      \phi_{-k}^{0}c_{k}^{0}+\phi_{-k}^{-1}c_{k}^{-1}+
      \phi^{L}_{-k}c_{k}^{L})\nonumber\\
&+& \sum_{k}(S_{-k}^{i,-1}\gamma_{k}^{i,-1}
    +S_{-k}^{i,1}\gamma_{k}^{i,1}+
    \sum_{i<j}\phi_{-k}^{ij}c_{k}^{ij})\nonumber\\
&+& \sum_{k,l}[-2ic_{-k}^{1}c_{-l}^{-1}b_{k+l}^{0}-
    ic_{-k}^{1}c_{-l}^{0}b_{k+l}^{1}+
    ic_{-k}^{-1}c_{-l}^{0}b_{k+l}^{-1}-\nonumber\\
& & (k+l)c_{-k}^{1}c_{-l}^{L}b_{k+l}^{1}
    -(k-l)c_{-k}^{-1}c_{-l}^{L}b_{k+l}^{-1}
    -kc_{-k}^{0}c_{-l}^{L}b_{k+l}^{0}-\nonumber\\
& & \half (k-l)c_{-k}^{L}c_{-l}^{L}b_{k+l}^{L}
    +(\frac{k}{2}-l)c_{-k}^{L}\gamma_{-l}^{i,-1}\beta_{k+l}^{i,-1}-
    \frac{1}{4} \gamma_{-k}^{i,-1}
\gamma_{-l}^{i,-1}b_{k+l}^{-1}+\nonumber\\
& & ic_{-k}^{1}\gamma_{-l}^{i,-1}\beta_{k+l}^{i,1}+
    \frac{i}{2}c_{-k}^{0}\gamma_{-l}^{i,-1}\beta_{k+l}^{i,-1}-
    ic_{-k}^{-1}\gamma_{-l}^{i,1}\beta_{k+l}^{i,-1}-\nonumber\\
& & (\frac{k}{2}+l)c_{-k}^{L}\gamma_{-l}^{i,1}\beta_{k+l}^{i,1}-
    \frac{i}{2}c_{-k}^{0}\gamma_{-l}^{i,1}\beta_{k+l}^{i,1}-
    \frac{1}{2} \gamma_{-k}^{i,-1}
\gamma_{-l}^{i,1}b_{k+l}^{0}-\nonumber\\
& & \frac{1}{4} \gamma_{-k}^{i,1}\gamma_{-l}^{i,1}b_{k+l}^{1}+
    \frac{i}{2} \gamma_{-k}^{i,-1}\gamma_{-l}^{j,1}b_{k+l}^{ij}-
    \frac{i}{2} \gamma_{-k}^{i,-1}\gamma_{-l}^{j,1}b_{k+l}^{ji}+\nonumber\\
& & lc_{-k}^{L}c_{-l}^{ij}b_{k+l}^{ij}-
    \half c_{-k}^{ij}c_{-l}^{j\alpha}b_{k+l}^{i\alpha}+
    \half c_{-k}^{ij}c_{-l}^{\alpha i}b_{k+l}^{\alpha j}+\nonumber\\
& & \half c_{-k}^{ij}c_{-l}^{i\alpha}b_{k+l}^{\alpha j}-
    \half c_{-k}^{ij}c_{-l}^{\alpha j}b_{k+l}^{i\alpha}-
    \half \gamma_{-k}^{i,-1}c_{-l}^{i\alpha}\beta_{k+l}^{\alpha,-1}+\nonumber\\
& & \half \gamma_{-k}^{i,-1}c_{-l}^{\alpha i}\beta_{k+l}^{\alpha,-1}-
    \half \gamma_{-k}^{i,1}c_{-l}^{i\alpha}\beta_{k+l}^{\alpha,1}+
    \half \gamma_{-k}^{i,1}c_{-l}^{\alpha i}\beta_{k+l}^{\alpha,1}].
\eee

We can check the nilpotency of $\cal{Q}$ with the use of the extended
 constraints.
Due to (\ref{tildephi}) and (\ref{Q3}) we will have
\be
\tilde{\phi}^{-1}_{m}
  &=&:\!\phi^{-1}_{m}\! :+ \sum_{-\infty}^{+\infty}:
      [2ic^{1}_{-k}b^{0}_{m+k}+ic^{0}_{-k}b^{-1}_{m+k}-
      (m-k)c^{L}_{-k}b^{-1}_{m+k}-\nonumber\\
  & & i\gamma^{i,1}_{-k}\beta^{i,-1}_{m+k}]:\label{exc1}  \\
\tilde{\phi}^{L}_{m}
  &=& :\! \phi^{L}_{m}\! :+\sum_{-\infty}^{+\infty}:
      [(k+m)c^{1}_{-k}b^{1}_{k+m}+(k-m)c^{-1}_{-k}b^{-1}_{m+k}+\nonumber \\
  & & kc^{0}_{-k}b^{0}_{m+k}+(k-m)c^{L}_{-k}b^{L}_{m+k}+
      kc^{ij}_{-k}b^{ij}_{m+k}+\nonumber\\
  & & (\frac{m}{2}-k)\gamma^{i,-1}_{-k}\beta^{i,-1}_{m+k}-
      (\frac{m}{2}+k)\gamma^{i,1}_{-k}\beta^{i,1}_{m+k}]:-\alpha_{L}
\delta_{m}\label{exc4}\\
\tilde{S}^{i,-1}_{m}
  &=& :\! S^{i,-1}_{m}\! :+ \sum_{-\infty}^{+\infty}:
      [(\frac{k}{2}-m)c^{L}_{-k}\beta^{i,-1}_{m+k}
      -\half \gamma^{i,-1}_{-k}b^{-1}_{m+k}-
      \frac{1}{2}c^{i\alpha}_{-k}\beta^{\alpha,-1}_{m+k}+\nonumber\\
  & & \frac{1}{2}c^{\alpha i}_{-k}\beta^{\alpha,-1}_{m+k}+
      ic^{1}_{-k}\beta^{i,1}_{m+k}+
      \frac{i}{2}c^{0}_{-k}\beta^{i,-1}_{m+k}-\nonumber\\
  & & \half \gamma^{i,1}_{-k}b^{0}_{m+k}+
      \frac{i}{2} \gamma^{j,1}_{-k}b^{ij}_{m+k}-
      \frac{i}{2} \gamma^{j,1}_{-k}b^{ji}_{m+k}]:\label{excS-1}\\
\tilde{\phi}^{0}_{m}
  &=& :\!\phi^{0}_{m}\! :+ \sum_{-\infty}^{+\infty}:
      [ic^{1}_{-k}b^{1}_{m+k}-ic^{-1}_{-k}b^{-1}_{m+k}-mc^{L}_{-k}b^{0}_{m+k}-
\nonumber\\
  & & \frac{i}{2} \gamma^{i,1}_{-k} \beta^{i,1}_{m+k}+
      \frac{i}{2} \gamma^{i,-1}_{-k} \beta^{i,-1}_{m+k}]:
      -\alpha_{0}\delta_{m}\label{excn2} \\
\tilde{\phi}^{1}_{m}
  &=& :\!\phi^{1}_{m}\! :+ \sum_{-\infty}^{+\infty}:
      [-2ic^{-1}_{-k}b^{0}_{m+k}-ic^{0}_{-k}b^{1}_{m+k}-
      (m+k)c^{L}_{-k}b^{1}_{m+k}+\nonumber\\
  & & i\gamma^{i,-1}_{-k}\beta^{i,1}_{m+k}]:\label{exc-1}  \\
\tilde{S}^{i,1}_{m}
  &=& :\! S^{i,1}_{m}\! :+ \sum_{-\infty}^{+\infty}:
      [-(\frac{k}{2}+m)c^{L}_{-k}\beta^{i,1}_{m+k}
      -\half \gamma^{i,1}_{-k}b^{1}_{m+k}-
      \frac{1}{2}c^{i\alpha}_{-k}\beta^{\alpha,1}_{m+k}+\nonumber\\
  & & \frac{1}{2}c^{\alpha i}_{-k}\beta^{\alpha,1}_{m+k}-
      ic^{-1}_{-k}\beta^{i,-1}_{m+k}-
      \frac{i}{2}c^{0}_{-k}\beta^{i,1}_{m+k}-\nonumber\\
  & & \half \gamma^{i,-1}_{-k}b^{0}_{m+k}-
      \frac{i}{2} \gamma^{j,-1}_{-k}b^{ij}_{m+k}+
      \frac{i}{2} \gamma^{j,-1}_{-k}b^{ji}_{m+k}]:\label{excS1}\\
\tilde{\phi}^{ij}_{m}
  &=& :\!\phi^{ij}_{m}\! :+ \sum_{-\infty}^{+\infty}:
      [-mc^{L}_{-k}b^{ij}_{m+k}+\half c^{i\alpha}_{-k}b^{\alpha j}_{m+k}+
      \half c^{i\alpha}_{-k}b^{j\alpha}_{m+k}-\nonumber\\
  & & \half c^{\alpha j}_{-k}b^{i\alpha}_{m+k}-
      \half c^{\alpha j}_{-k}b^{\alpha i}_{m+k}-
      \half \gamma^{i,-1}_{-k}\beta^{j,-1}_{m+k}+\nonumber\\
  & & \half \gamma^{j,-1}_{-k}\beta^{i,-1}_{m+k}-
      \half \gamma^{i,1}_{-k}\beta^{j,1}_{m+k}+
      \half \gamma^{j,1}_{-k}\beta^{i,1}_{m+k}]:.\label{excij}
\ee
They satisfy the same algebra as the original constraints. This means
 that (\ref{Q})
\be\label{Q4}
{\cal Q}^{2}&=&\tilde{d}_{1} \sum_{m}
\frac{m^{3}}{2}c^{L}_{m}c^{L}_{-m}\nonumber\\
 &+&\tilde{d}_{2}\sum_{m} \frac{m}{2}c^{L}_{m}c^{L}_{-m}\nonumber\\
 &+&\tilde{d}_{3}
\sum_{m} (-i\frac{m}{2}c^{0}_{m}c^{0}_{-m}+m^{2}c^{0}_{m}c^{L}_{-m}+2i
mc^{1}_{m}c^{-1}_{-m}-m\gamma^{i,-1}_{m}\gamma^{i,1}_{-m}) \nonumber\\
 &+&\tilde{d}_{4}
\sum_{m}(mc^{0}_{m}c^{L}_{-m}+2i
c^{1}_{m}c^{-1}_{-m}+\half \gamma^{i,-1}_{m}\gamma^{i,1}_{-m}).
\ee

The only thing that remains now is to calculate the values of the
 constants $\tilde{d_{f}}=
1,\ldots,4$ for the vacuum and ordering introduced  introduced in the
 previous section. In
this case the condition $ p^{A}_{m}|0\rangle =0 \quad \forall A$ together
 with the
requirement that the BRST charge (\ref{Q3}) gives the following
 consistency conditions
$\forall m$

\bbe{FullVacCond2}
  p^{A}_m|0\rangle
 =c^{1}_m|0\rangle=\gamma^{i,1}_m|0\rangle=b^{-1}_m|0\rangle =
\beta^{i,-1}_m|0\rangle=0\nonumber\\
  \langle 0|x^{A}_m  =\langle 0|c^{-1}_m= \langle 0|\gamma^{i,-1}_m=
 \langle 0|b^{1}_m
=\langle 0|\beta^{i,1}_m=0.\nonumber
\eee

We  find the values of $\tilde{d_{3}}$ and $\tilde{d_{4}}$ by calculating  the
expectation value of the commutator $\left\langle
0\right|[\tilde{\phi}^{0}_{m},\tilde{\phi}^{L}_{-m}]\left|
  0\right\rangle $.
Indeed we have
\be\label{fcon2}
&\left\langle 0\right|[\tilde{\phi}^{0}_{m},\tilde{\phi}^{L}_{-m}]\left|
  0\right\rangle =m\left\langle 0\right|\tilde{\phi}^{0}_{0}\left|
  0\right\rangle +\tilde{d}_{3}m^{2}+\tilde{d}_{4}m&\nonumber\\
  &\Rightarrow 0=m\alpha_{0}+\tilde{d}_{3}m^{2}
  +\tilde{d}_{4}m=0&\nonumber\\
  &\Rightarrow \tilde{d}_{3}=0,\qquad
  \tilde{d}_{4}=-\alpha_{0}.&\nonumber
\ee
But from (\ref{excn2}) for a Hermitian BRST charge $\cal{Q}$ we will have
\be\label{ano}
\alpha_{0}\equiv \left\langle 0\right|\tilde{\phi}^{0}_{0}\left|
  0\right\rangle =-\frac{i}{4}[d-2+2N]\lim_{K\to +\infty}\left(
 \sum_{k=-K}^{+K}1
\right).
\ee

The BRST charge is nilpotent when all the constants $\tilde{d_{f}}=
1,\ldots,4$ are equal to 0. In particular we must have
\be\label{final}
\tilde{d}_{4}=0 \Rightarrow d=2-2N.
\ee
This last relation shows us that we cannot have a positive critical
 dimension for $N\neq 0$
irrespective of how $\psi^{i}_{A}$ acts on vacuum. Note also that
 (\ref{final}) gives
$d=2$ for $N=0$ as obtained in \cite{us}.
\raggedbottom

In this article we carried the techniques developed in \cite{us}
for the conformal string,
to the $T\rightarrow 0$ limit of the spinning string. Faced with
the anomaly (\ref{ano}),
it would be interesting to see if we could use the philosophy of
\cite{BigT} which led to a
"topological state space" result. The natural generalization in
our case would be to
require that ${\cal Q}^{2}=0$ only on physical states. This is
a topic of future
investigations. Another interesting problem is to see if it is
possible to apply the same
techniques to the space-time superstring case and to investigate
if this theory is
equivalent to the one presented here.

\bigskip
\begin{flushleft}
{\bf Acknowledgements:}I would like to  thank Ulf Lindstr\"om and
Rikard von Unge for useful
comments and fruitful discussions.
\bigskip
\end{flushleft}

\eject

\end{document}